# Independent Co-Assurance using the Safety-Security Assurance Framework (SSAF): A Bayesian Belief Network Implementation for IEC 61508 and Common Criteria

**Nikita Johnson[1], Youcef Gheraibia and Tim Kelly**

University of York, UK

**Abstract**  For modern safety-critical systems we aim to simultaneously maintain safety whilst taking advantage of the benefits of system interconnectedness and faster communications. Many standards have recognised and responded to the serious security implications of making these connections between systems that have traditionally been closed. In addition, there have been several advances in developing techniques to combine the two attributes, however, the problem of integrated assurance remains. What is missing is a systematic approach to reasoning about alignment. In this paper, the Safety-Security Assurance Framework (SSAF) is presented as a candidate solution. SSAF is a two part framework based on the concept of independent co-assurance (i.e. allowing separate assurance processes, but enabling the timely exchange of correct and necessary information). To demonstrate SSAF's application, a case study is given using requirements from widely-adopted standards (IEC 61508 and Common Criteria) and a Bayesian Belief Network. With a clear understanding of the trade-offs and the interactions, it is possible to create better models for alignment and therefore improve safety-security co-assurance.

## 1 Introduction

In systems engineering the tension between system safety (the desire to protect people from harm) and cyber security (the desire to protect assets of a system) has increased in recent years. This is primarily due to the increased size and interconnectivity of modern systems. In order to maximise productivity and capability we allow systems to communicate with each other and share their services. This interconnectedness presents greater security risk as there is a larger attack surface and many more ways for 'things to go wrong'. This threatens to undermine the goals of an entire system, including safety goals. From an assurance perspective, it is therefore no longer acceptable to assume complete separation of safety and security risk. Nor is it acceptable to treat that risk solely in the comfortable isolation of each domain's[2] practices, knowledge, and culture.

---

[1] Corresponding author nlj500 <at> york.ac.uk

In an attempt to address the issue of isolated practice, many solutions have been created by extending existing safety techniques. These are partial solutions at best because their focus is predominantly on safety and much of the security information is discarded. Many of these techniques have previously been critically examined (Johnson & Kelly, 2019a) and found insufficient for through-life co-assurance.

There are *technical approaches* that aim to combine risk concepts across domains; *organisational structures* that allow for better communication between experts; and also legal and *regulatory initiatives* to align[3] safety and security at national and international levels. However, these changes do not address all the concerns that arise from bringing the two attributes together, therefore issues with misalignment remain.

This paper presents a candidate solution that enables and supports full alignment of safety and security – the Safety Security Assurance Framework (SSAF). At the core of the Framework is the concept of *independent co-assurance* and *synchronisation points*. This allows for some separation to be maintained in order to make the most of differing expertise, knowledge and practice, whilst ensuring that the right people get the right information at the right time.

This paper is laid out in three parts. Part 1: Section 2 attempts to characterise the differences between safety and security. Part 2: Section 3 and 4 present the Safety Security Assurance Framework (SSAF) and apply it to a case study using a Bayesian approach and requirements found in standards. Finally, Part 3: Sections 5, 6 and 7 examine the rationale of the decisions made during the case study, discuss related work, and provide a conclusion.

## 2    Characterising Challenges in Safety and Security Assurance

Due to the similarities between safety and security assurance, namely that both processes are concerned with reduction of risk and prevention of loss, it is tempting to equate the two notions of risk and have one representative (quantitative) value. Whilst this is certainly feasible, as demonstrated by many techniques that combine risk analysis into a single methodology, these methodologies often disregard the fact that there exist differences that make the attributes incommensurable[4]. This may be one of the reasons that none of these methodologies has been widely adopted for co-assuring safety-critical systems.

---

[2] Here *domain* refers to technical area (safety or security) rather than application domain.

[3] The term *alignment* will be used interchangeably with co-assurance because of its widespread use in industry. It is worth noting that co-assurance (argument and process) is achieved through aligning safety and security.

[4] Definition: not able to be judged by the same standards.

Two characterisations of safety and security assurance will be discussed in this section with the aim of making their differences clear. The emphasis is on what makes co-assurance difficult, and the requisite qualities of an effective alignment solution. The first characterisation will look at the contribution of security concerns to safety risk; the second will use widely-adopted standards to establish differences between the attributes.

## 2.1 Effects of Security on Safety Risk

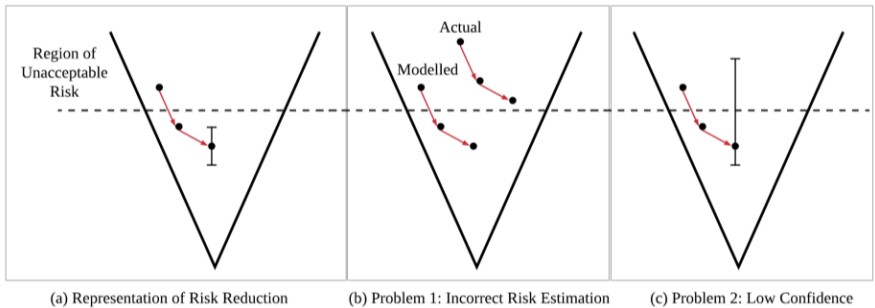

(a) Representation of Risk Reduction  (b) Problem 1: Incorrect Risk Estimation  (c) Problem 2: Low Confidence

**Fig. 1.** ALARP Representation of the Problems of Uncertainty for Safety Risk

In the UK, the Health and Safety at Work Act 1974 (HSE, 1974) states it is the duty of employers to ensure the safety of its employees "so far as is reasonably practicable". This philosophy is often enacted as the ALARP principle: safety risk should be As Low As Reasonably Practicable. Depicted in Figure 1(a) is the ALARP "carrot diagram". The idea is to identify the level of a particular risk, then systematically reduce that risk until it is ALARP. It is possible to reduce risk in one of three ways: *i.* Designing it out of a system, *ii.* Engineering in controls, or *iii.* Having procedural mitigations.

Alongside the risk value is a window of variation which is analogous to a statistical confidence interval; this represents the uncertainty in the estimation of risk. Several factors affect this interval such as the competence of the practitioners, the rigour of their processes, the limitations of the tools they use, *etc.* Safety is concerned with the higher portion of this interval, and the potential for variance into the unacceptable risk region. Thus, it is often a requirement by regulators for a confidence argument to be provided with the safety risk argument or safety case.

Figure 1(b) shows the first problem of safety-security alignment. Practitioners and engineers might follow an ALARP process and use their expert judgement to estimate the level of a particular risk; however due to the presence of an intelligent and motivated adversary the level of risk might be substantially higher in reality. Therefore, models and artefacts used to support a safety case are inaccurate and the safety argument is fundamentally undermined. There are ways that this can be minimised, for example verifying estimates made at design time against operational data, however this is not always feasible.

Figure 1(c) shows the second problem for the safety-security interaction: there may exist an estimation of risk, but the level of uncertainty may be high due to security concerns. This could be the result of socio-technical factors, such as inadequate processes, or the judgement of a practitioner with insufficient training.

Whilst the underlying reason for these two co-assurance problems is the uncertainty introduced by security concerns, there are different treatments of uncertainty. Most existing technical approaches focus solely on the uncertainty introduced in Problem 1 above, *i.e.* they attempt to improve the accuracy of risk level by considering security sources of risk, but do not consider the implications of other assurance factors. This leads to:

**Solution Criterion 1 (SC1)** A solution for co-assurance *must* facilitate reasoning about the diverse ways in which uncertainty is introduced into assurance, and how it is handled on multiple levels of abstraction.

## 2.2 Safety and Security Assurance Characteristics

In (Alexander et al. 2004), characteristics of Systems-of-Systems are identified with the aim of prompting analysts to proactively consider likely failure modes in relation to the characteristics; thus, another way to gain knowledge about risk, besides learning from accidents, is created.

In a similar fashion, in this section, the safety and security characteristics for which there is conflict are identified from three of the most commonly used standards across both domains. The aim of identifying the characteristics is to understand the ways in which safety and security can diverge, and so proactively consider ways to prevent these *co-assurance failures*.

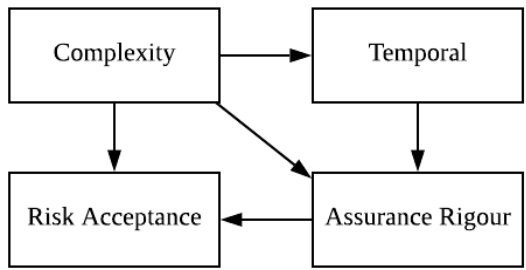

**Fig. 2.** Characteristics of Co-Assurance Factors

The source of safety assurance characteristics is IEC 61508 (International Electrotechnical Commission, 2010). For security, both ISO 27005 (International Or-ganization for Standardization, 2011) and Common Criteria (International Organization for Standardization, 2017) were used. Figure 2 shows the shared characteristics of safety and security assurance.

*Risk Acceptance* – This describes the attributes' attitude to risk. First and foremost, safety is about preventing death and injury so naturally it is very conservative and risk averse; made clear from the prescriptive and conservative safety requirements. This is in stark contrast to security, where acceptable risk is a lot more strategic; in most systems (that are not safety-critical) there is an element of gameplay[5] in an attempt to balance security risk against potential benefits. This flexibility is reflected in the standards, where unlike safety standards, no universal notion of acceptable risk is given. It is worth noting that as security incidents begin to have a greater impact on society, regulatory authorities and lawmakers are attempting to define acceptable risk for security so that if it is not maintained, punitive action can be taken for those responsible.

*Assurance Rigour* – Both attributes' assurance processes aim for order, reasoned arguments supported by evidence, predictable behaviours, and repeatable outcomes, *etc*. Due to safety's risk aversion, the operational environment of safety-critical systems is often defined and strictly bounded, thereby making it easier to achieve higher levels of assurance rigour at system level. For security however, the intelligent adversary alone means that there exists higher levels of uncertainty, which makes it much more difficult to achieve predictable outcomes. The approach is therefore to achieve assurance rigour at lower levels of the system (*e.g.* component level) with clearly defined assumptions to test against.

*Temporal* – This is the time required to perform assurance activities and the expected rate of change. For safety, the conservative risk stance and the need to provide detailed argumentation about acceptable risk takes time. It is also expected that once a system has been certified[6], the safety case remains valid unless there is a major change to the system or its usage. In contrast, the security assurance environment is more agile and fast-paced. This is reflected in the "classes" approach to requirements in the standards that aims to protect against groups of vulnerabilities, so that change can be responded to more effectively – such as when a new vulnerability is discovered. Security arguments need to be robust to change at a faster rate than safety.

*Complexity* – Both safety and security are emergent, but they handle different types of complexity in different ways. For example, complexity due to number of components is not an issue in and of itself for safety[7]; what is important is that complexity does not interfere with the assurance factors for safety. In contrast, redundancy and diversity creates more work for security, as there are a greater number of potential attack options to consider. This has a significant impact on the time resource available and the level of assurance rigour that can be achieved.

---

[5] From Game Theory where rational decision-makers maximise rewards within given bounds.

[6] "certified" used in the broadest sense. Based on the assumption that *all* safety-related sectors have some form of formal safety acceptance before use that does not necessary involve a regulator.

[7] For example, if a system is built with an infinite number of components whose behaviour is formally understood, then we can create a mathematical model of this infinitely large system, and argue safety.

These are the four characteristics where the most significant alignment failures are likely to occur. Having defined these, it is now possible to find mitigations to prevent the specific mismatches. Therefore:

**Solution Criterion 2 (SC2)** A solution for co-assurance *must* address the alignment trade-offs with respect to each of the assurance characteristics (risk acceptance, assurance rigour, temporal and complexity).

It is clear that the co-assurance solution criteria stated in this section are not inherently technical. Although there will most certainly be an element of technical analysis required for alignment, a solution that considers only this aspect and not the wider context would be a partial solution at best. What is needed is an approach that has the capability to address both the technical risk alignment and the socio-technical assurance factors.

## 3 The Safety-Security Assurance Framework

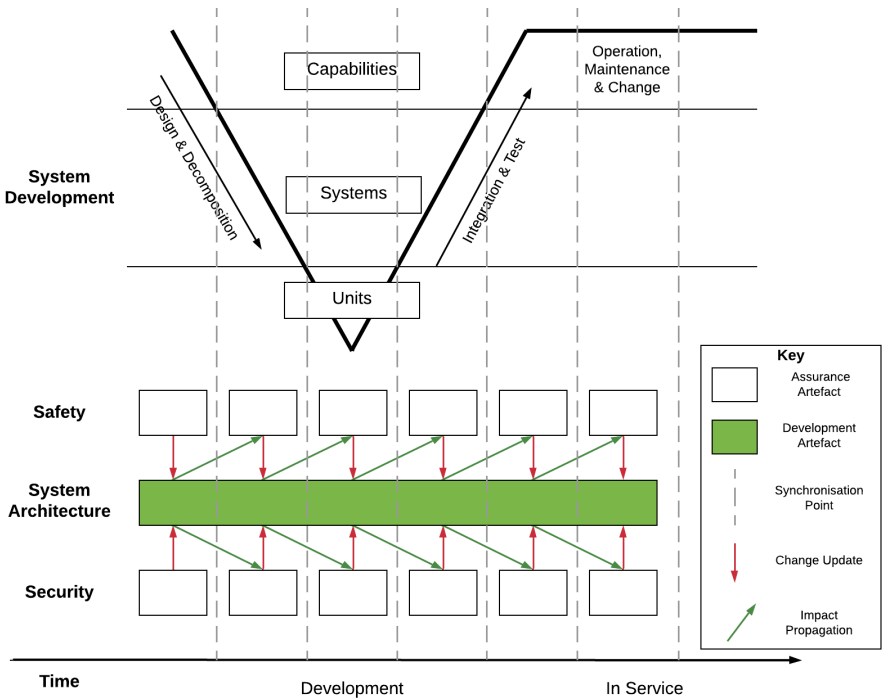

**Fig. 3.** SSAF: Independent Co-assurance and Synchronisation Points

The Safety-Security Assurance Framework (SSAF) is a solution for co-assurance that addresses the criteria (SC1 and SC2) stipulated in the challenges section. It consists of a process and model for systematically reasoning about the alignment of system safety and cyber security throughout the life of a system. SSAF is built on the new paradigm of *independent co-assurance* – that is, keeping the disciplines and expertise separate but having key synchronisation points where required information is exchanged across the discipline boundaries *i.e.* "the right information is given to the right people at the right time". Figure 3 illustrates this concept during system development and deployment. Note that SSAF was developed with the assumption of a model-based design, however that does not preclude its use on non-model-based systems. More work is required to establish the synchronisation (sync) points in this case.

SSAF is comprised of two models – the *Technical Risk Model* allows for the communication of risk and impact across disciplines; and the *Socio-Technical Model* which recognises that co-assurance is an inherently human activity that involves many judgements at different levels that could constrain the technical solution to alignment.

## 3.1 SSAF Technical Risk Model (TRM)

The first SSAF model – the **Technical Risk Model** has three major parts:
1. An *ontology* for cross discipline communication
2. A *5-step process* for creating synchronisation points and links
3. A *causal model and patterns* for different conditions and their relationships

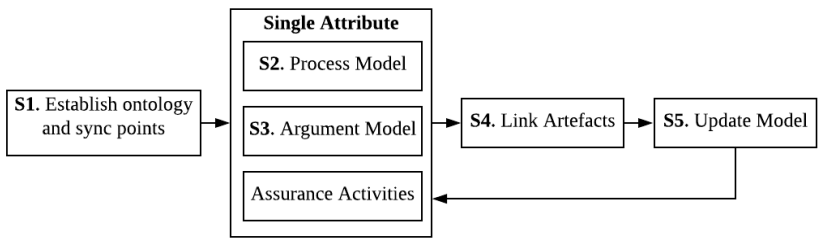

**Fig. 4.** SSAF TRM Process for Synchronisation

The SSAF TRM processes in Figure 4 consists of five steps. Steps 1, 4 and 5 are where the unique contribution of the TRM lies. Steps 2 and 3 are assumed to be the standard best practice in each of the domains[8].

---

[8] In (Johnson & Kelly, 2019b) a worked example of the application of the SSAF TRM process to a wearable medical device is given.

At the core of making independent co-assurance, and therefore SSAF, work is establishing sync points and understanding the causal relationships between conditions in safety and security. Many standards have started to include synchronisation points where safety and security must communicate as an integral part of their processes *e.g.* ARP 4754A (SAE International, 2010) and DO-326A (RTCA, 2014) for aerospace, ISO 14971 (International Organization for Standardization, 2007, p. 200) and AAMI TIR 57 (Association for the Advancement of Medical Instrumentation, 2016) for medical devices, *etc.* However, it can be argued that these are the absolute minimum needed for alignment. In order to add more sync points and improve alignment, it is important to understand what information needs to be exchanged and when.

Figure 5 shows the (partial) TRM causal meta-model that enables relationships between conditions across domains to be modelled. By systematically understanding relationship between conditions and the synchronisation points where they occur – patterns of interaction can be created. This helps to address the problem of inaccurate risk estimation in each domain, but particularly for safety[9].

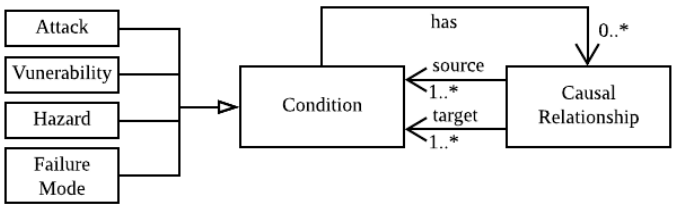

**Fig. 5.** SSAF TRM Causal Relationships represented in UML

## 3.2 SSAF Socio-Technical Model (STM)

The second part of SSAF is the Socio-Technical Model (STM). Whilst modelling causal relationships for technical risk alignment is a very powerful concept, it is insufficient when considering that there may be a discrepancy between the *modelled* condition and how it occurs in actuality. STM helps to minimise the effect of this problem by understanding the sources of uncertainty that affect confidence[10;11].

Figure 6 shows the STM model of interactions. Unlike the TRM it does not consider causal relationships, rather, it uses another type of assurance influence – the confidence relationship. The idea of explicitly modelling confidence is not unique to SSAF STM. The Motor Industry Software Reliability Association have included the requirement to reason about organisational and individual factors that affect technical risk (MISRA, 2019). SSAF STM identifies two types of confidence.

---

[9] Addresses problem 1: Inaccurate Risk Estimation from Section 2.1

[10] Addresses problem 2: High Uncertainty with regards to Risk Estimation from Section 2.1

[11] In (Johnson & Kelly, 2019c) a worked example about how to use the STM is provided.

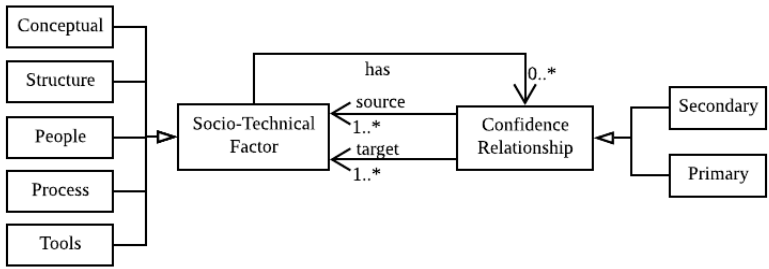

**Fig. 6.** SSAF STM Confidence Relationships

*Primary confidence* describes all the factors that could affect the technical risk argument directly, examples of this include the analysis approach being used, and the competence of practitioners.

*Secondary confidence* describes all the factors that may influence primary confidence. This includes organisational structures, engineering processes, individual cognitive biases, *etc*.

The STM has roots in the socio-technical systems domain. The categories of assurance factors are based on Bostrom and Heinan's model of socio-technical systems design (Bostrom & Heinen, 1977). The categories are *Structure*, *Process People* and *Tools*. An additional category, *Conceptual*, was added to the model. It is orthogonal to the other types of assurance factor. For safety-security alignment communication of mental models is one of the biggest factors affecting assurance. The communication of concepts was not adequately represented in any of the other four categories. By explicitly representing points where uncertainty can be propagated across domain boundaries, we can start to reason about *why* it is not a problem, and still have confidence in the risk argument. Without articulating these relationships it's almost impossible to do this.

## *3.3 Assurance surface*

Another useful concept introduced by the SSAF model is that of the assurance surface. The underlying concept is borrowed from the security domain - *attack surface i.e.* all the vectors that an attacker might exploit to launch an attack on a system (Howard, Pincus, & Wing, 2005). The *assurance surface* by analogy is all the ways in which uncertainty can be propagated across domains. The concept necessitates an important shift in thinking for co-assurance. It implies that there are multiple ways in which uncertainty can be propagated, and therefore it is highly unlikely that any one technical approach to integration will address all uncertainty propagation. Instead we must think in terms of *assurance coverage*, and use the best possible combination of approaches and argumentation to minimise uncertainty propagation across domains. Figure 7 shows the relation between SSAF TRM and STM to existing safety-critical system development. More detail about the tiers is provided in (Johnson & Kelly, 2019c).

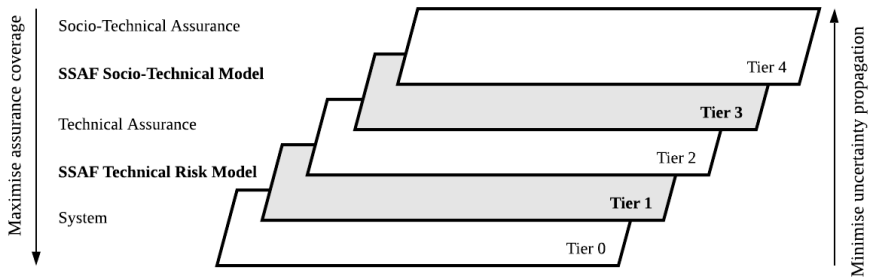

**Fig. 7.** SSAF Assurance Surface Concept

## 4   Case Study: SSAF TRM Bayesian Approach to Standards

Thus far, SSAF independent co-assurance has been presented as a potential solution to safety-security alignment problems. The reasoning behind why the approach of explicitly modelling relationships is the best way forward has been provided. However, still lacking is the means by which to achieve this.

That is the purpose of this case study. The motivation is twofold – first, to provide empirical evidence for the SSAF models and methodology. Secondly, to show the creation of a re-useable alignment pattern. To maximise applicability the case study, requirements from two of the most commonly used single-domain standards were selected: IEC 61508 and Common Criteria.

### 4.1 IEC 61508 and Common Criteria Brief

IEC 61508 is arguably the most widely adopted safety standard. It has been adapted to multiple domains including healthcare, rail, automotive, aerospace, and nuclear. It consists of seven parts that define the safety process for a system. It is the software design and development (software architecture design) requirements found in Table A.2, p48 IEC 61508-3:2010 that were selected for this case study.

Common Criteria is a widely adopted security standard that has been adopted across many types of systems in many domains, including some that are safety-critical. It consists of three parts. For this case study, functional requirements from Common Criteria Part 2 were selected.

## 4.2 'Link and Sync' Methodology

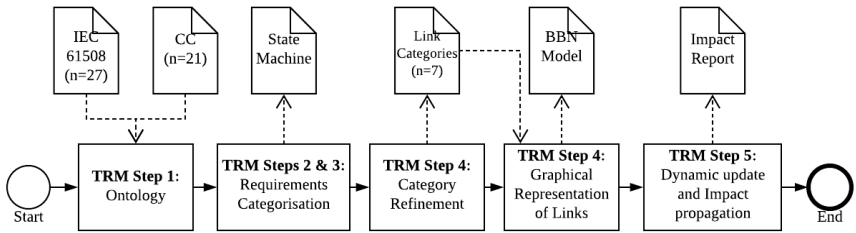

**Fig. 8.** BPMN[12] Model of Methodology for Linking Safety and Security Requirements

The main contribution of SSAF TRM is the explicit modelling of and reasoning about the causal relationships that exist at different synchronisation points. This is encapsulated in Steps 1, 4 and 5 of the SSAF TRM Process. The objective of this case study is to demonstrate *how* this process functions, emulate how industrial system requirements could be linked, and show how the links could be implemented on a project. Figure 8 shows the process steps followed for the case study.

**Step 1 – Ontology.** Using 27 functional design requirements from IEC 61508 (found in Part 3 Annex A Table A.2) and 21 functional requirements from Common Criteria Part 2 – commonalities and general categories were identified.

**Steps 2 & 3 – Requirements Categorisation.** These steps were performed independently within each domain, with respect to either safety or security. The ontology and categories established in Step 1 were used to categorise the requirements according to type. In addition, a state machine was created to explain the impact on safety in the absence of a safety argument (further detail in Section 5.2).

**Step 4a – Category Refinement.** Once the requirements had been through initial categorisation, the categories were jointly refined further which resulted in 7 types of requirements. These were mapped to four states in a state machine that showed which requirements were violated. The four states were St0 None, St1 Resource & Timing Violated, St2 Failure Behaviour Violated, St3 Communication Violated.

**Step 4b – Graphical Representation.** Using the refined categories, requirements from safety and security which were in the same categories were linked to each other. These links were then modelled as a Bayesian Belief Network (BBN).

**Step 5 – Dynamic Update and Impact Propagation.** The leaf nodes of the BBN are the security classes of requirements from Common Criteria. A practitioner provides details if a security requirement class has been "violated" or not. The BBN then outputs the probabilities of being in state St1, St2 or St3.

---

[12] Business Process Model and Notation – process modelling notation

## *4.3 Results*

Figure 9 shows the state machine that was output from TRM Steps 2 and 3. It consists of four states. S0 where no safety requirements have been violated, and three other states where at least one safety requirement from the IEC 61508 set was violated. Transitions occur according to the type of safety requirement that has not been satisfied[13], for example not satisfying requirement "13a Guaranteed maximum time" would transition to state S1. To return to S0 that violation would need to be resolved. The states were formed by grouping the seven requirements types in groups which were highly cohesive, *i.e.* {Resource Use and Timing}, {Failure Behaviour, Failure Detection, Recovery}, and {Communication and Trust}.

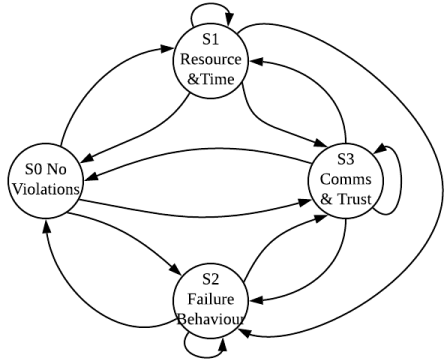

**Fig. 9.** State Machine for Safety Requirements Violation

Figures 10(a) and 10(b) show the model of the causal links that were established during the linking process in TRM Step 4. Figure 10(a) provides a summary conceptual model to communicate the content and structure of the BBN. Figure 10(b) shows the real-world implementation of the BBN in the GeNIe modelling tool[14].

The leaf nodes of the BBN are the requirements classes taken from Common Criteria. The driving concept that makes this model successful, is the idea that if any of the security requirements in that class are violated, it can be input into the BBN leaf nodes. The impact then propagates through the classes and related safety requirements to the safety output *i.e.* the impact report which is the probability of being in a particular safety state.

---

[13] It does not add to the analysis in this case to distinguish between "requirements violation" and "not satisfying a requirement". They are viewed as equivalent, however it might be necessary to make the distinction in operational systems where violation during operation may carry more severe consequences than requirements not being met pre-deployment at design.

[14] Tool description can be found at: https://www.bayesfusion.com/genie/

As knowledge is contained in the state machine about how to transition back to a state where no safety requirements have been violated, it is now possible for a safety practitioner to take the output impact report from the BBN, and use that to determine the state, then resolve the issue more efficiently without needing to know specific information about the security requirements.

This model would be most useful during operation where security violations can occur at a fast rate. However, the model has some utility during the requirements phase to reason about impact – similar to sensitivity analysis.

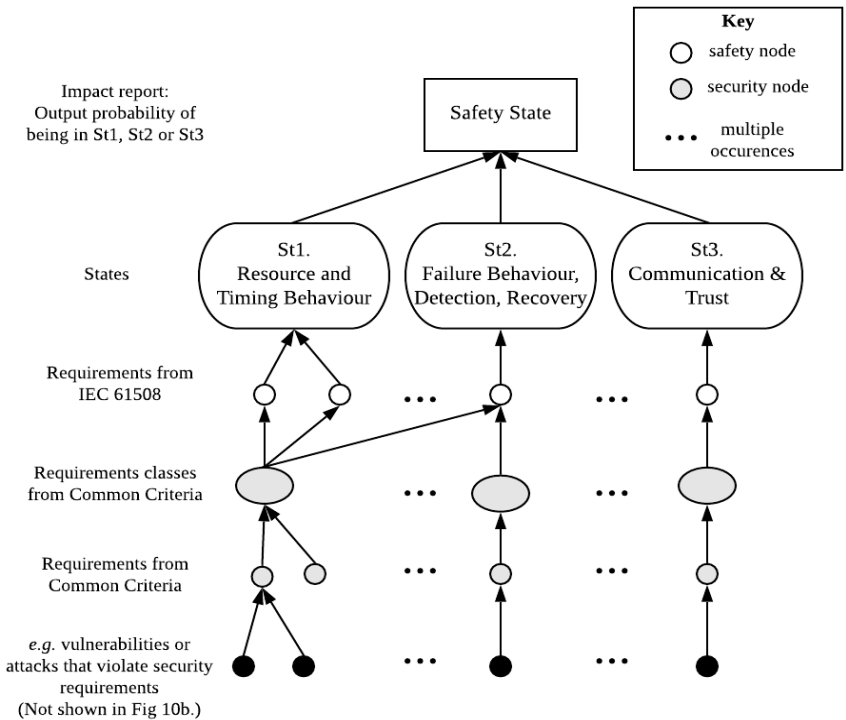

**Fig. 10(a).** Conceptual Model of BBN Links between Safety and Security

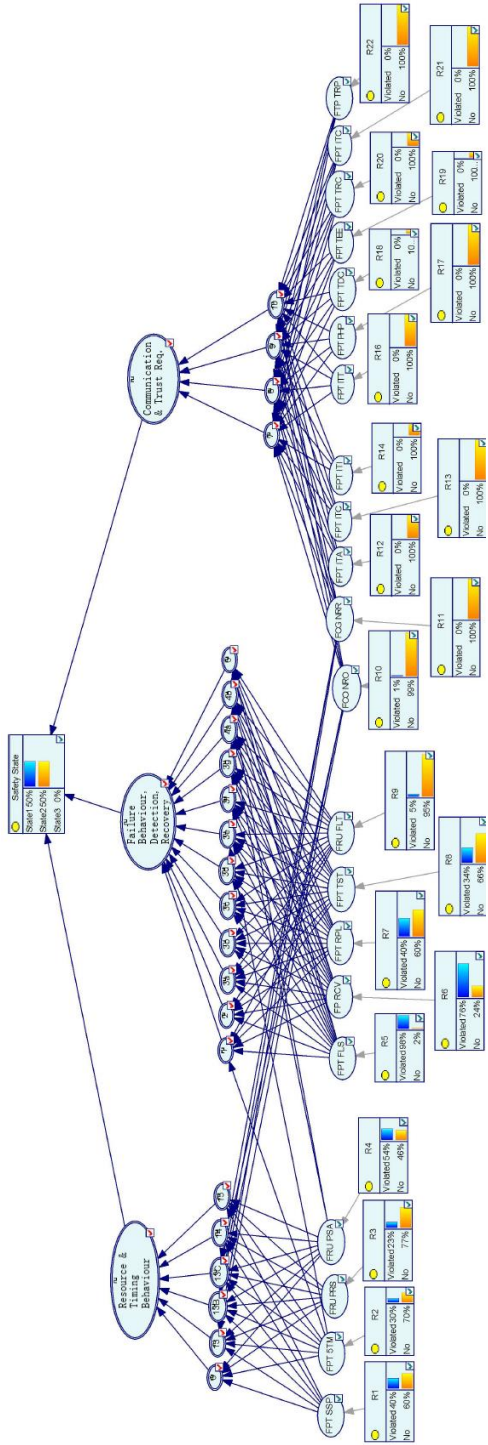

**Fig. 10(b).** GeNIe Implementation of the BBN from Fig. 10(a)

## 5 Discussion

The quality of this implementation of SSAF is dependent on the quality of the links *i.e.* between the safety requirements from IEC 61508 and security requirements from Common Criteria. The links were determined by sorting them in to cohesive groups. Table 1 shows an example of the requirements from both safety and security for Resource Use and Timing. If performed on an industrial project, the group categories could be decided beforehand, practitioners could classify the artefacts in each domain separately, subsequently link tables can be created.

**Table 1.** Example Grouping for Resource Use and Timing Requirements

| Domain | ID | Requirement |
|---|---|---|
| Safety | A2.6 | Dynamic reconfiguration |
| | A2.13a | Guaranteed maximum time |
| | A2.13b | Time-triggered architecture |
| | A2.13c | Maximum response to events |
| | A2.14 | Static resource allocation |
| | A2.15 | Static synchronisation of access |
| Security | FPT_SSP | State synchrony protocol |
| | FPT_STM | Time stamps |
| | FRU_PRS | Priority of service |
| | FRU_RSA | Resource allocation |

### *5.1 Deciding the Causal Relationships*

Deciding on group categories is a non-trivial task. Most unified methodologies such as security-informed fault trees usually specify the syntax of how artefacts should be linked, but not the semantics, *e.g.* linking the top event of an attack tree to the base event of a fault tree. This SSAF Case Study implementation goes some way to answering the question about semantics of causal links. In this case, expert judgement, experience and concept cohesion were used to make the groupings.

It would have been possible to create links with less complex reasoning behind them, such as linking all safety and security requirements per component; however, the aim of co-assurance is to argue alignment using these links, therefore a more structured and strategic approach is needed for link creation. Preferably, an approach that can be examined, contested and repeated if necessary.

Grounded Theory (GT) research methodology (Glaser & Strauss, 2017) was used to address this problem because its function is to build connections that are grounded in data. Ordinarily, GT is used in social science and humanities research. It was selected for application in this case study because it structures the thinking of the person performing it. It also has several distinct phases such as initial sampling, populating, memoing, constant comparison, *etc.* that can be documented and reviewed. This explicit documentation of reasoning could be used as evidence to support an alignment argument. Another advantage of using a GT approach is that existing connections can be extended or new ones created. This is especially important when considering multiple, complicated notions of causality that are present for safety and security. Having a causal model that theory of interaction that can be expanded is essential.

## 5.2 Action After Determining the Impact

In the TRM process described in previous work (Johnson & Kelly, 2019b) there exists the assumption that the argument structures for each attribute are known. In addition, there is the assumption that the artefacts (*e.g.* analysis models used for evidence) are linked to the argument (*e.g.* safety case) and the TRM model. So when a change occurs, impact can be traced from the TRM model to the claims in the argument. However, modelling the argument structures for Common Criteria and IEC 61508 was beyond the scope of this case study which is concerned mainly with the creation of causal links.

Instead, a state machine was presented as a way to understand the security impact on safety; *i.e.* by construction, the states communicate to the safety practitioner which types of safety requirement have been violated. This allows safety practitioners and decision makers to respond to change more effectively because they are not required to reason about security requirements to understand impact. Knowledge about particular states and how to transition is encapsulated in the model. This approach enables resources to be applied proportionally to the impact. For example, from a safety perspective, moving to a state where an availability-related requirement has been violated (S2) is probably of more concern than if a confidentiality-related requirement has been violated (S3).

There are, of course, a few limitations to using the state machine for the purposes of determining impact. The first is the assumption that the transitions modelled are possible and accurate, *i.e.* once a safety requirement has been violated then a suitable and timely resolution can be found to transition back to state S0 where there are no violations; this is unlikely to be true in all cases. However, even if transitions are not possible it is still important to capture the reasoning and assumptions in a systematic way.

Another limitation is the simplicity of the model. Only four states were modelled for comprehensibility, but many more states could be captured with many more complex transitions. States could be included to represent multiple violations, par-tial violations, *etc.* - this would risk a possible state explosion that would be counterproductive to the aim of using the model, *i.e.* for practitioners to understand impact and what to do next.

Although there are limitations with this approach to handling impact, the state machine is understandable and helps practitioners to respond proportionally.

### 5.3 Socio-Technical Considerations

The Case Study was a controlled application of SSAF TRM whose focus was on establishing causal links and propagating technical risk. Thus, it is quite difficult to evaluate the socio-technical factors as they would have occurred in an industrial project such as evaluating preparatory tasks for the alignment meetings, *etc.*

SSAF STM provides a structured model to help practitioners reason about the socio-technical factors that would affect technical risk. Table 2 shows the kinds of claims that might be examined. Some of these, labelled primary, are claims that would affect confidence of the technical risk argument. All other factors affect the socio-technical factors that impact risk, those are labelled secondary.

**Table 2.** Example SSAF Socio-Technical Claims

| STM Factor | Confidence | Claim |
|---|---|---|
| People | Primary | Practitioners are sufficiently competent to perform the methodologies. |
| Process | Secondary | Timing is bounded for information exchange and issue resolution |
| Structure | Secondary | The responsibility and authority to manage safety-security interactions has been designated |
| Tools | Primary | BBN is sufficient for the purpose of modelling the interactions between requirements |

In the SSAF Case Study example, the alignment argument with primary and secondary confidence claims would need to be provided to show that the BBN was fit-for-purpose. A possible rebuttal for this particular instance, is that BBNs do not include a time dimension, therefore another complementary methodology might be used to align temporal factors.

## 6  Evaluation and Related Work

In Section 2, two criteria were identified for an alignment solution:

**Solution Criterion 1 (SC1)** A solution for co-assurance *must* facilitate reasoning about the diverse ways in which uncertainty is introduced into assurance, and how it is handled on multiple levels of abstraction.

**Solution Criterion 2 (SC2)** A solution for co-assurance *must* address the alignment trade-offs with respect to each of the assurance characteristics (risk acceptance, assurance rigour, temporal and complexity).

The SSAF Case Study presented here was limited to finding causal links between requirements for risk impact propagation. It is important to note that this is just one type of information exchange between safety and security, at one very particular point in the development lifecycle. The advantage of using the SSAF approach is that it does not limit practitioners to using just this one synchronisation point and methodology for alignment. Instead, it allows for multiple complementary approaches to be used. SC1 is therefore satisfied because SSAF provides a structure for reasoning at multiple synchronisation points throughout the life of a system. SC2 is much more difficult to demonstrate without a full industrial application and evaluation because the assurance factors are context dependent. However, even without this context, SSAF provides the mechanisms needed to address each of the factors.

Other than some emerging standards that include cross-domain considerations, SSAF is the only framework that allows for bi-directional impact propagation. . Even then, the standards have a limited number of synchronisation points which would reflect minimum best practice at the time that the standard was written. SSAF is flexible enough to support practitioners if alignment approach requires many synchronisation points. SSAF also provides the structure to argue about the alignment of both safety and security from both a technical and socio-technical perspective. Another advantage, is that its models and output can provide evidence for an explicit co-assurance argument which could support other certification and accreditation activities.

Making the causal links and reasoning behind the co-assurance argument explicit has many advantages as discussed earlier in this section, but it also presents new questions such as whether or not those links or reasoning are correct, what the relationship is between links on different levels of abstraction and how to review the quality of the links with resources that could be spent engineering the system rather than on assurance.

## 6.1 Related Work

The case study showed one implementation of one causal model at one particular synchronisation point. This model will clearly not be universally applicable even though it is useful. It is possible that models such as this one could be used as the basis for alignment patterns which would allow practitioners and engineers to reuse the knowledge.

There are many existing techniques for safety-security co-assurance and co-engineering. If the underlying causal model was made explicit, then more patterns could be created, thus creating a catalogue of possible alignment strategies. Some patterns could also be used as a requirement of standards. This is likely to be a much more practical approach to alignment than stating a synchronisation point but not saying how information should be exchanged, or providing one unified technique that is limited to only partial analysis of co-assurance factors.

Table 3 (discussed in Johnson & Kelly (2019b)) shows methodologies for alignment, and the possible causal links they represent:

**Table 3.** Causal Relationship Examples

| Condition | | Causal Relationship | |
|---|---|---|---|
| **Source** | **Target** | **Label** | **Methodology[15]** |
| Vulnerabilities | Failure | causes | FFA |
| Vulnerabilities | Hazards | contribute to | SAHARA, DDA, UML, FTA |
| Safety Effect | Attack | motivates | ADT |
| Threat | Hazard | safety impact | Standard |
| Threat | Safety Requirements | Influence | STPA-Sec, STPA-SafeSec |
| Safety Requirements | Security Requirements | trade-off | ATAM |
| Security Requirements | Safety Requirements | trade-off | ATAM |
| Security Controls | Safety Requirements | conflict with | ad-hoc |

# 7    Conclusion

The Safety-Security Assurance Framework was presented as a candidate solution to the existing alignment problem between system safety and cyber security. SSAF is based on a new paradigm of independent co-assurance which allows for separate domains, expertise, ways of working, *etc.* but requires that predetermined synchronisation points are established where information is exchanged. Multiple methodologies can be used at these synchronisation points, commensurate with the needs for alignment.

---

[15] Full references and some discussion about these methodologies is given in Johnson & Kelly 2019b. The methodologies refer to Functional Failure Analysis (FFA), Security-Aware Hazard Analysis and Risk Assessment (SAHARA; HARA from ISO 26262 Part 3), Dependability Deviation Analysis (DDA), Unified Modelling Language (UML), Fault Tree Analysis (FTA), Attack Defense Tree (ADT), Systems Theorectic Process Analysis – Sec/SafeSec (STPA-Sec and STPA-SafeSec), and Achitecture Tradeoff Analysis Method (ATAM).

Much like the role of a systems integrator, it is likely that a new role will be created to manage the co-assurance argument and artefacts, and to ensure that the necessary activities occur; otherwise the co-assurance goals that are not covered by either safety or security goals will be overlooked.

The vision for the future of safety and security co-assurance is that the knowledge and practice for alignment is captured and modelled explicitly so that it can be examined, reasoned about, contested and reused. SSAF provides the structure to make that possible.

**Acknowledgments**   Research and development of SSAF supported by the University of York, the Assuring Autonomy International Programme (AAIP), and BAE Systems. Research Award Ref: EPSRC iCASE 1515047.